# GAIN-SWITCHED VCSEL AS A QUANTUM ENTROPY SOURCE: THE PROBLEM OF QUANTUM AND CLASSICAL NOISE

[1,2]**R.A. Shakhovoy**[✉], [1]**E.I. Maksimova,**
[1]QRate, 100 Novaya str., Skolkovo, Russia
[2]NTI Center for Quantum Communications, National University of Science and Technology
[1]MISiS, 4 Leninsky prospekt, Moscow, Russia
[✉]r.shakhovoy @goqrate.com

**Abstract.** We consider the problem of quantum noise extraction from polarization swapping in a gain-switched VCSEL. The principle of operation of a quantum random number generator is based on the generation of laser pulses with one of two orthogonal polarization states, followed by digitization of polarization-resolved pulses with a comparator. At intensity values of laser pulses close to the threshold value of the comparator, the contribution of the classical noise of the photodetector will have a crucial role in making a decision on the choice of a logical zero or one. We show how to evaluate the contribution of classical noise and how to calculate the quantum reduction factor required for post-processing.

**Keywords:** quantum random number generators, vertical surface emitting laser, quantum noise extraction

**Funding:** This work was supported by Russian Science Foundation (grant no. 17-71-20146).

## Introduction

Random number generators (RNGs) play a primary role in modern cryptographic applications. Due to the development of quantum cryptography, a special place among RNGs occupy now quantum RNGs (QRNGs), which use various quantum sources of entropy. Over the past 15-20 years, a number of approaches have been proposed to obtain quantum randomness; however, optical QRNGs, which employ laser noise, have gained the most popularity. Laser noise can be associated with various effects, e.g., with temperature-related fluctuations of the laser cavity length or with pump fluctuations. However, at relatively high frequencies, laser noise is associated mainly with spontaneous emission occurring due to zero-point oscillations of an electromagnetic field, which have purely quantum nature and are generally considered to have the properties of white noise. Due to this, laser noise can be employed as a high-frequency source of quantum entropy.

The main difference between optical QRNGs based on laser noise lies in how the noise is measured. Thus, interference-based optical QRNGs use *phase* noise of laser radiation, which is converted into amplitude fluctuations in the interferometer and then is readily measured with conventional photodetectors. In lasers that do not have fixed polarization of light, one may also use fluctuations of the *polarization state* in addition to phase fluctuations. Such an approach can be used, e.g., in a vertical-cavity surface-emitting laser (VCSEL). A VCSEL-based QRNG employing spontaneous polarization switching was first described in [1], where the author demonstrated the random bit generation rate up to 2 Mbps. Recently, we discussed a simple optical scheme of a QRNG based on a gain-switched VCSEL, which allows generating the sequence of random "on-off" pulses at several gigahertz [2]. Experimental demonstration of theoretical calculations performed there have been published in [3]. In the present article, we consider the problem of quantum noise extraction from polarization swapping in a gain-switched VCSEL. We use the approach developed in [4], namely, we introduce for the QRNG under consideration the *quantum reduction factor* containing information on the amount of classical noise "falling" into the digitized random sequence due to fluctuations in the photodetector. We also describe how this classical noise can be filtered out with the post-processing procedure.

## Simulations

A simplified scheme of the proposed QRNG is shown in Fig. 1(a). A gain-switched VCSEL is driven by a high-frequency laser driver, which is, in turn, controlled by the computer or FPGA. Laser output is followed by the polarization filter (PF) that allows obtaining polarization-resolved optical pulses, which are converted into the electrical signal via a broadband photodetector (PD). Random bits are obtained by digitizing pulses with a comparator, whose threshold voltage is calculated in the FPGA, which also performs post-processing procedures including randomness extraction.

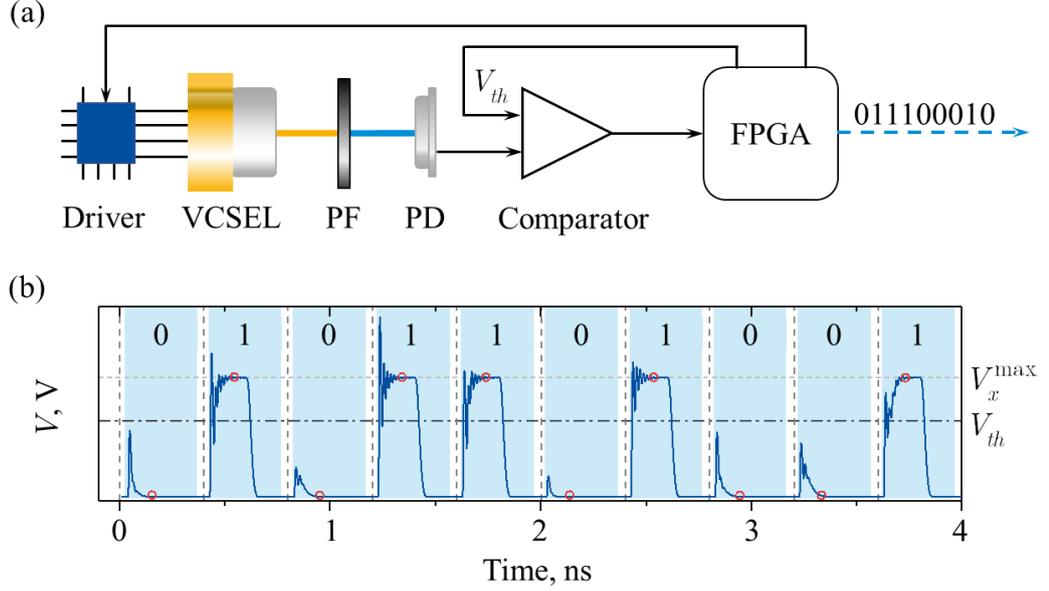

**Fig. 1.** (a) A simplified scheme of a VCSEL-based QRNG.
(b) Pulses at the comparator input and the results of the digitization

In Fig. 1(b), we simulated the digitization process of polarization-resolved laser pulses. It was assumed in simulations that the polarizer in Fig. 1(a) passes to the photodetector the $x$-linear polarization. Laser pulses were simulated with VCSEL rate equations given in [1]; the calculated signal was then processed with a low-pass digital filter (30 GHz bandpass) to simulate the finite bandwidth of the photodetector. The level of the comparator threshold ($V_{th}$) is shown by the dash-dotted line in Fig. 1(b); red circles correspond to the moments of the comparator latch actuation. The result of digitization ('0'-s or '1'-s) is shown in the corresponding frames (each time frame is shown by the blue rectangle).

Generally, a laser pulse at the VCSEL output contains both polarization components ($x$ and $y$), such that polarization state of a given pulse can be referred to as "quasi-elliptical". Relative contribution of orthogonal components is a random quantity; however, it depends on the width of the pulse and the rate of relaxation processes (transients). To demonstrate this, we calculated probability density function (PDF) of the normalized integral signal $S_x$ at three different repetition rates (Fig. 2(a)). In the ideal case, we would get two peaks at the values $S_x = 0$ and $S_x = 1$, which means that all optical power goes into one particular linear polarization ($y$ and $x$ respectively). However, due to the finiteness of transients, $S_x$ could take intermediate values between 0 and 1. The influence of transients becomes more prominent when decreasing the pulse width, which is clearly seen in Fig. 2(a), where the area under the PDF curve in the middle of the histogram is increased when increasing the pulse repetition rate from 2.5 to 7 GHz. Polarization-resolved laser pulses ($x$-pulses in our case) that fall into this intermediate region are the most affected by classical (non-quantum) noises of the photodetector; therefore, '0'-s and '1'-s resulted from digitization of such pulses can be considered as "untrusted" bits. The proportion of these bits can be thought of as a quantum reduction factor $\tilde{\Gamma}$, whose value determines how much the raw random sequence should be "compressed" using the randomness extractor. We propose the following formula to find $\tilde{\Gamma}$:

$$\tilde{\Gamma} = \frac{1}{H_\infty (1-P)} \quad (1)$$

where $H_\infty$ is the min-entropy of the raw random sequence, and $P$ is the probability to obtain the pulse with the $S_x$ value inside some "window" around the middle of the probability distribution, whose width is proportional to the relative r.m.s. value of the photodetector noise $\sigma$ ("measured" in terms of the normalized value $S_x$).

We also calculated the dependence of $\tilde{\Gamma}$ on the photodetector's noise $\sigma$ at different pulse repetition rates (see Fig. 2(b)). One can see that the reduction factor $\tilde{\Gamma}$ grows when increasing the pulse repetition rate and begins to grow faster with increasing $\sigma$. It means that it does not make much sense to increase the repetition rate of laser pulses if the photodetector is quite noisy.

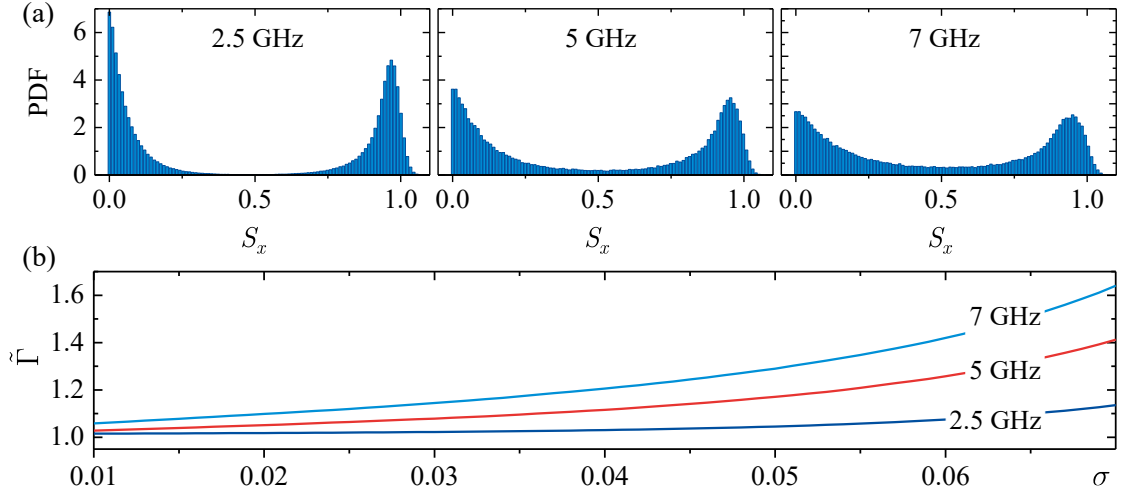

**Fig. 2.** Probability densities of the normalized integral signal $S_x$ (a) and dependences of the reduction factor $\tilde{\Gamma}$ on the photodetector's noise $\sigma$ (b) at different pulse repetition rates.

### Post-processing

The digitized random sequence in the proposed scheme must be subjected to the randomness extraction procedure with the reduction factor $\tilde{\Gamma}$, defined by (1). We may, however, use a deterministic extractor, e.g., the von Neumann extractor [5], which extracts randomness regardless the value of the reduction factor. The von Neumann extractor discards repeated bits in a sequence and replaces the two-bit words '01' and '10' with bits '0' and '1', respectively. Unfortunately, this extractor reduces the length of a sequence by at least 4 times, which is not very efficient. Therefore, one generally uses instead a seeded extractor. In cryptographic applications, an extractor with a seed is generally implemented in the form of 2-universal hash functions, whose efficiency is guaranteed by the leftover hash lemma [6]. A common way to implement 2-universal hashing is to multiply the input raw sequence by a random binary matrix [7]. Without loss of generality, one may always use for these purposes random Boolean Toeplitz matrices, which allow significantly saving the seed length. In our case, the randomness extractor is then divided into three steps:
1) For a "raw" binary sequence of length $n$, determine the length of the output sequence $m$ by the formula: $m = n/\tilde{\Gamma}$.
2) Generate the Toeplitz matrix using the "seed" of length $m+n-1$ bits.
3) Multiply the Toeplitz matrix by the raw sequence. This yields the resulting random sequence.

It is important to discuss the method of obtaining the seed. By default, it is assumed that the seed is obtained from a strong source of entropy, i.e., one that allows getting truly random bits. If the RNG being developed is not a strong source of entropy, then an additional source of entropy must be used. We propose, however, the following algorithm to obtain the seed. When switching-on the QRNG, the system buffers a raw random sequence of a given (relatively small) length. Then, this sequence is subjected to a deterministic extractor, e.g., the von Neumann extractor. The random sequence obtained after the extractor can be now used as a seed in hashing algorithms.

"Equipped" with such a procedure, the QRNG under consideration is an autonomous source of entropy that does not need an additional entropy source, i.e., the device can operate even in the absence of a pre-memorized random sequence.

One of the common ways to test the quality of randomness, is to perform statistical tests, e.g., NIST tests [8]. Unfortunately, we did not have an access to real (obtained in the experiment) random numbers; however, we have a fairly detailed theoretical model, which may be used to follow the whole route the laser noise "travels" from spontaneous emission to the sequence of random bits. For this, we simulated $10^6$ laser pulses similar to those shown in Fig. 1(b). To "digitize" them, the energy of each pulse (area under the pulse) was calculated and compared with a certain threshold energy. The obtained random bits were then grouped into $k$-bit words (we put $k=8$), which we denote as $x[i]$. The sequence of $x[i]$ were processed with a second-order FIR filter according to the following formula: $y[i] = \mod(x[i] + 2x[i-1] + x[i-2], 2^k)$, where each $y[i]$ is an $i$-th output word. The obtained data were then concatenate to a ("filtered") random bit string. After filtering, we processed the data with the randomness extractor described above. Finally, we performed NIST tests with random bits; the results of the test are summarized in Fig. 3. As one can see, the obtained random sequence successfully passed all the test.

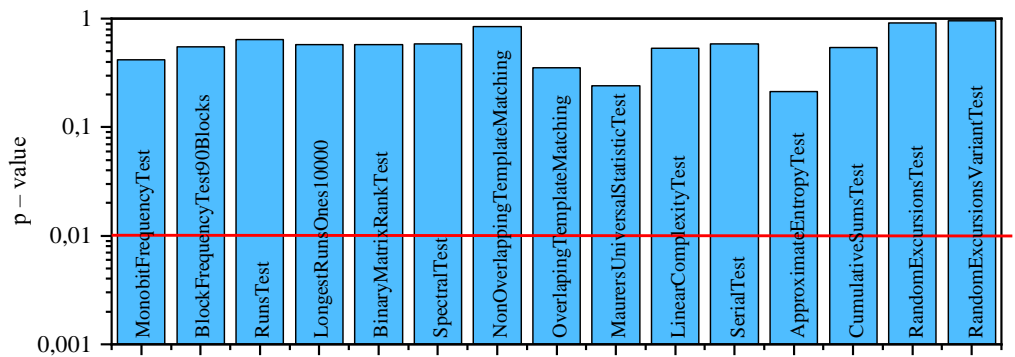

**Fig. 3.** NIST statistical test result.

## Conclusion

We propose here an approach for quantum noise extraction from polarization swapping in a gain-switched VCSEL and proposed a simple method to get a seed for hashing the raw random sequence without an additional entropy source. The discussed algorithms allow developing a fully autonomous QRNG with proven "quantumness" of generated random bits.

## Acknowledgments

Authors are grateful to Vladimir Meshkov for valuable comments.